\documentclass{ws-procs9x6}

\newcommand{\lsim}{\raisebox{-0.13cm}{~\shortstack{$<$ \\[-0.07cm] $\sim$}}~}
\newcommand{\gsim}{\raisebox{-0.13cm}{~\shortstack{$>$ \\[-0.07cm] $\sim$}}~}

\begin{document}

\title{SCENARIOS OF GRAVITINO DARK MATTER AND 
THEIR COSMOLOGICAL AND PARTICLE PHYSICS IMPLICATIONS}

\author{G. MOULTAKA$^*$}

\address{Laboratoire de Physique Th\'eorique et Astroparticules \\
{\sl UMR5207--CNRS}, Universit\'e Montpellier II \\
Place E. Bataillon, F--34095 Montpellier Cedex 5, France\\
$^*$E-mail: Gilbert.Moultaka@lpta.univ-montp2.fr\\
www.univ-montp2.fr}

%

\begin{abstract}
I report on some scenarios where the gravitino is the dark matter and
the supersymmetry breaking mediated by a gauge sector.   
\end{abstract}


\bodymatter

\section{Introduction}\label{sec:introduction}
Providing a viable particle candidate for the non-baryonic dark matter (DM) in the
Universe has become one of the main test requirements for model building beyond
the standard model (SM) of particle physics. It is well-known  
(see for instance \cite{Jungman:1995df}) that an electrically neutral  particle,
weakly interacting with the primordial plasma and with a mass of order the 
electroweak scale ($\lsim {\cal O}(1 \, \rm TeV)$ ) would
 have today a relic density $\Omega \simeq 1$, provided 
 it is stable or sufficiently long-lived, thus putting $\Omega h^2$ 
 in the ballpark 
 of the WMAP results \cite{Komatsu:2008hk}. It should then not come so much 
 as a surprise that most scenarios beyond the SM can provide potential solutions to the 
 dark matter mystery, even less take it as an indication for their particular 
 physical relevance.
 Rather, one should keep in mind that {\sl i)} the abovementioned estimate
 of $\Omega$ assumes a simple thermal history of the early Universe
 {\sl ii)} only a few classes of the proposed scenarios beyond the SM are theoretically 
 framed in what was their motivation in the first place, i.e.
  solve the shortcomings of the standard model of particle physics.
 
 We take hereafter point {\sl ii)} as our guiding principle and address the
 question of dark matter from that point of view. We will then see that the 
 assumption of point {\sl i)} does not always apply in typical parameter 
 space regions of the scenarios under consideration.

\section{Supersymmetric extensions} The supersymmetric (SUSY) extensions of the
SM are among the most fashionable examples of {\sl ii)}, including (at least)
 the ingredients of the minimal supersymmetric standard model (MSSM). Yet not  all
of them provide a DM candidate in the configuration {\sl i)}. However, 
supersymmetry breaking being the trigger of the electroweak symmetry breaking, 
it is justified to study on the same footing the physical consequences
of different SUSY  breaking and mediation scenarios. For instance, in gravity
mediated SUSY breaking scenarios \cite{Chamseddine:1982jx},
\cite{Barbieri:1982eh}, \cite{Hall:1983iz} it is natural to expect the
gravitino mass $m_{\tilde G}$ to be of order the electroweak scale, thus
leaving room for a massive neutral weakly interacting particle such as a
Neutralino to be the lightest SUSY particle. If stable, such a particle would 
perfectly fit point {\sl i)} and provide a very good DM candidate. This
tremendously studied scenario since the work of \cite{Goldberg:1983nd},
\cite{Ellis:1983ew}, as natural as it may look, still relies on two crucial
assumptions: the lightest susy particle (LSP) is not electrically charged
(typically such as the tau  slepton ($\tilde{\tau}$) ) {\sl and} there is a 
residual R-parity guaranteeing the stability (or at least a sufficiently  long
lifetime) of the lightest Neutralino. Theoretically, these two assumptions are
not necessarily favored
\footnote{apart from the requirement itself of tailoring a DM candidate!} 
since they can strongly depend on  the actual  dynamical
mechanism underlying SUSY breaking, which is still poorly understood. 
An alternative
option which has attracted much attention in recent years is to take the
gravitino as the LSP, another logical possibility within the context of gravity
mediated supersymmetry breaking, see for instance \cite{Olive:2008uf}.

In this presentation we put the focus on a different kind of scenarios
where the SUSY breaking  and its mediation to the supersymmetric standard model
is realized through some gauge interactions
\cite{Fayet:1978qc,
Dine:1981za, Dimopoulos:1981au,Dine:1981gu, Dine:1982qj, Dine:1982zb,
AlvarezGaume:1981wy, Nappi:1982hm, Dimopoulos:1982gm}, 
\cite{Dine:1993yw, Dine:1994vc,Dine:1995ag}. The models originating from
this class of gauge mediated susy breaking (GMSB) scenarios are 
phenomenologically as
compelling as the gravity mediated ones, and have 
similar theoretical uncertainties. An important difference however 
is that here the gravitino is 
very light and {\sl necessarily} the LSP, rather than this being a possibility 
among others. As usual, one can concoct exceptions. (See for instance \cite{Shirai:2008qt}
for a model where the gravitino is not the LSP, bringing the case back to the 
Neutralino DM configurations.) 
We stick however to the generic GMSB, assuming
$m_{\tilde G} \lsim {\cal O}(1 \; \rm GeV)$;
 in this case the DM issue is somewhat tricky,
and in particular does not quite fit point {\sl i)}. 

\section{The phenomenological GMSB}\label{sec:GMSB} 
\begin{figure}[t]
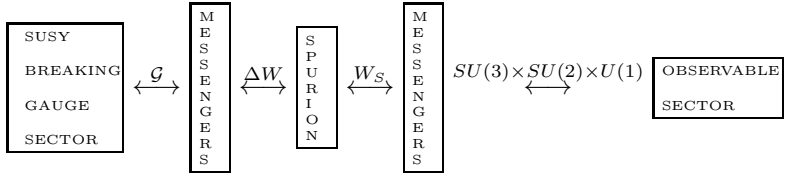

\fbox{ \parbox{1.2cm}{{\tiny SUSY

BREAKING
GAUGE
SECTOR}}} $\stackrel{\small{{\cal G}}}{\longleftrightarrow}$ \fbox{\parbox{.3cm}{\tiny{M

E

S

S

E

N 

G

E

R

S}}} $\stackrel{\Delta W}{\longleftrightarrow}$ \fbox{\parbox{.3cm}{{\tiny 
S

P

U

R

I

O

N

}}} $\stackrel{W_S}{\longleftrightarrow}$ \fbox{\parbox{.3cm}{\tiny{M

E

S

S

E

N 

G

E

R

S}}}
$\stackrel{{\tiny{SU(3) \times SU(2) \times U(1)}}}{\longleftrightarrow}$ \fbox{\parbox{1.5cm}{{\tiny OBSERVABLE

SECTOR}}}
\caption{generic structure of GMSB model sectors.}
\label{fig:fig1}
\end{figure}

We recall hereafter the main
phenomenological ingredients of the gauge mediated susy breaking models based
on the assumption that the leading contribution to the dynamical susy breaking
is originating from some strongly coupled gauge sector (SBGS), screened
(often dubbed 'hidden' or 'secluded') from the
visible MSSM sector by some intermediate non-gauge interactions
\cite{Dine:1993yw, Dine:1994vc,Dine:1995ag}. The various sectors are schematized as boxes in
\fref{fig:fig1} and the  interactions among them indicated by the arrows. 
The two messenger sectors are formed of matter chiral superfields 
$\hat{\Phi}_M, \hat{\overline{\Phi}}_M$
and $\hat{\phi}_m,...$, 
 having rather similar status; they have gauge interactions respectively with 
 the visible MSSM sector and the hidden susy breaking sector, and non gauge 
interactions, through the superpotential, with an intermediate spurionic 
chiral superfield $\hat{S}$.    
$\hat{\Phi}_M$  and  $\hat{\overline{\Phi}}_M$ have  quark-like or lepton-like 
charges  under $SU(3)_c \times SU(2)_L \times U(1)_Y$,
$\Phi \sim (3, 1, -\frac{1}{3}) \; \mbox{or} \; (1, 2, \frac{1}{2})$,
$\overline{\Phi}_M \sim (\bar{3}, 1, \frac{1}{3}) \; \mbox{or} \; 
(1, 2, -\frac{1}{2})$. To preserve gauge coupling unification these fields
are usually put into larger gauge group multiplets,
{\sl e.g.} $\mathbf{5} + \bar{\mathbf{5}}$ or 
$\mathbf{10} + \overline{\mathbf{10}}$ of SU(5)${}_{{}_{\rm GUT}}$, 
and $\mathbf{16} + \overline{\mathbf{16}}$ of SO(10)${}_{{}_{\rm GUT}}$.
The other messengers $\hat{\phi}_m$ are charged under some gauge group
${\cal G}$ through which they feel the properties of the susy breaking 
(secluded) sector.\footnote{We do not enter here the fascinating question  of 
susy breaking through non-perturbative gauge 
interaction phenomena supposed to occur in the latter sector, which were studied since 
the early eighties (see \cite{Nilles:1983ge,  Giudice:1998bp} for reviews)
and rejuvinated recently
\cite{Intriligator:2006dd}.}
Furthermore, the messenger fields on both sides are assumed to interact 
only indirectly via $\hat{S}$ through the  
superpotential
$W \supset   W_S + \Delta W(\hat{S},\hat{\phi}_i) + W_{\rm MSSM}$
where, $W_{\rm MSSM}$ is the visible sector superpotential, 
$W_S = \kappa \hat{S}\hat{\Phi}_M\hat{\overline{\Phi}}_M + 
\frac{\lambda}{3} \hat{S}^3$
and $\hat{S}$ is neutral under all the gauge groups involved.

In the sequel we will be mainly interested in the three sectors on the
right-thand side of \fref{fig:fig1}. The conditions under which it is justified
to ignore the effects of the two other sectors in the early Universe will be 
touched upon in \sref{sec:cosmo}. From the phenomenological point of view, all 
we need to assume here about these two sectors is that they cooperate
to give non-zero vacuum expectation values (vev), $\langle S \rangle$ and
$\langle F_S \rangle$, respectively to the scalar and F-term components
of $\hat{S}$. This gives a supersymmetric mass 
$M_{f}=  \kappa   \langle S \rangle \equiv M_X$, 
to the (Dirac) fermion component as well as SUSY breaking mass 
spectrum $M_{s_\pm} = M_X(1 \pm \frac{\kappa  \langle F_S \rangle}{M_X^2})^{1/2}$
to the mass eigenstates of the scalar components of the 
$\hat{\Phi}_M,  \hat{\overline{\Phi}}_M$ fields. The amount of susy
breaking transmitted to the messenger/MSSM sectors, $\langle F_S \rangle$,  
is in general only a fraction of the total amount of the SUSY breaking in 
the SBGS which we denote $\langle F_{\rm TOT} \rangle$. The fermionic component
$\psi_S$ of $\hat{S}$ will then carry a fraction of the goldstino in the form
$\psi_S = \frac{\langle F_S \rangle}{ \langle F_{\rm TOT} \rangle} {\tilde G} +
\dots$. It  will thus contribute to the coupling of the massive 
gravitino to matter via its spin-$1 \over 2$ component. 
Last but not least, the SUSY breaking is communicated to the visible sector
through the gauge interactions of the messengers, leading to gaugino and
scalar soft masses
in the MSSM respectively at the $1-$ and  $2-$loop 
levels in the form  
$M_i\sim \left({\alpha_i\over 4\pi}\right){<F_S>\over
M_X}\,$ and $\tilde{m}_a^2 \sim \left({\alpha_a\over 4\pi}\right)^2
\left({<F_S>\over M_X}\right)^2 $  (where $\alpha_{i,a}$ denote the SM gauge
couplings or combinations thereof, and we have dropped   for simplicity
detailed flavor, messenger number and loop dependent coefficients).
Assuming a typical grand unified group, the full MSSM and messenger spectrum
and couplings depend  uniquely on 
three continuous and one discrete parameters, namely $M_X$, 
${<F_S>\over M_X} (\equiv \Lambda)$, $\tan \beta$ (the ratio of the two
higgs doublet vevs), and $N_{\rm mess}$  
the number of quark-like/lepton-like  messenger multiplets of some GUT group.
\footnote{for simplicity we fix the couplings relevant to the interactions with the 
spurion as $\kappa, \lambda  \simeq 1$ in $W_S$, and choose the MSSM
$\mu$-parameter  to be 
positive; note also that the trilinear soft parameter $A_0$ is vanishing to leading
loop order.} 

Finally we note that the model defined so far possesses a discrete symmetry
implying the conservation of the number of messengers in each physical
process. An important consequence is that the lightest messenger particles (LMP)
with mass $M_{s_{-}}$
will be stable due to such a symmetry. As we will see in \sref{sec:cosmo}
such stable particles can have a dramatic cosmological effect. 
Furthermore, depending on the GUT group multiplets they belong to,
the mass degeneracy of these LMPs can be lifted by quantum corrections 
\cite{Dimopoulos:1996gy, Han:1997wn} leading to an LMP with very specific
quantum numbers; e.g. $\tilde{\nu}_L$-like or $\tilde{e}_L$-like, if in a 
$\mathbf{5} + \bar{\mathbf{5}}$ of SU(5), an electrically charged SU(2)${}_L$ 
singlet, if in a $\mathbf{10} + \overline{\mathbf{10}}$ of SU(5), 
and an MSSM singlet, if in a $\mathbf{16} + \overline{\mathbf{16}}$
of SO(10).

%

\section{Coupling to Supergravity}\label{sec:sugra} The gravitino being the
gauge field of local supersymmetry and the superpartner of the graviton, its
proper inclusion in the model of the previous section necessitates the coupling
of the latter to supergravity. The ensuing rich structure allows to determine 
fairly uniquely all the couplings of the gravitino to the  other states (such
that flat space-time GMSB models are retrieved in the limit of infinite Planck
mass) \cite{Wess:1992cp}. Coupling to supergravity has some  other benefits:
--the massless spin-$1 \over 2$ goldstino, originating from the spontaneous
SUSY  breaking in the SBGS, appears as a mixture of the fermions of all chiral
(resp. vector) supermultiplets whose F-terms (resp. D-terms)  develop vevs, and
ceases to be a physical state since it mixes automatically with the  massless 
spin-$3 \over 2$ gravitino giving a mass to the latter. This is the origin of
the relation between $\psi_S$ and $\tilde{G}$ noted in the previous section,
where $\langle F_{\rm TOT} \rangle$ includes all F-term and D-term  vevs  --the
requirement of an (almost)  vanishing cosmological constant, together with that of SUSY
breaking, leads to the general relation  $\langle F_{TOT} \rangle   \simeq
\sqrt{3} \; m_{\tilde{G}} \; m_{\rm Pl} $, where $ m_{\rm Pl}$ denotes the
reduced Planck mass ($\simeq 2.4 \times 10^{18}$GeV). Combined with the
qualitative   relation $G_F^{-1/2} \sim {\langle F_S \rangle \over M_X }$
between the electroweak scale and the two mass scales at one's  disposal in the
visible sector,  one finds $m_{\tilde{G}} \sim {G_F^{-1/2}} {M_X \over \sqrt{3}
m_{\rm Pl} k} $  (where $k \equiv  \langle F_S \rangle/ \langle F_{\rm TOT}
\rangle$), that is typically a very light gravitino in gauge mediated models since
$M_X \ll m_{\rm Pl}$ (as compared to gravity mediation where it becomes of 
order the electroweak scale with  $M_X \sim O(m_{\rm Pl})$) 
--since GMSB generates soft gaugino masses $M_i$, the spontaneous SUSY 
breaking is accompanied by a spontaneous breaking of a continuous R symmetry.
The latter leads to an R-axion which is phenomenologically problematic.
However, the requirement mentioned previously
for the cosmological constant is actually achieved through an additive constant
$W_0 =m_{\tilde{G}} m_{\rm Pl}^2$ to the superpotential $W$ of \sref{sec:GMSB}
(if there are no Planck scale vevs),
thus breaking explicitly the R symmetry and giving possibly very large masses 
to the R-axion \cite{Bagger:1994hh}   --including gravity suggests that
some discrete symmetries valid in flat space could be violated by
non-perturbative quantum gravitational effects, involving for instance
black hole physics \cite{Dolgov:1991wv
}. In particular,
the accidental messenger number conservation responsible for the stabilitly
of the LMP can be lost, leading to Planck suppressed decays into
MSSM particles through effective (non-)renormalizable messenger number
violating (but gauge invariant \cite{Krauss:1988zc}) operators. 
Such operators can appear either in the superpotential or in the K\"ahler
potential, the latter being further organized into holomorphic
or non-holomorphic in the fields; e.g., taking the SU(5) GUT particle content
and messengers in $\mathbf{5}_M + \bar{\mathbf{5}}_M$, one can have
$K_{\rm hol} \supset \mathbf{5}_{M} \bar{\mathbf{5}}_F, \,  
 {1 \over m_{Pl}} \times \bigl\{   \bar{\mathbf{5}}_M\bar{\mathbf{5}}_{F,H}\mathbf{10}_F\,,\,
\mathbf{5}_M\mathbf{10}_F\mathbf{10}_F \dots \bigr\}$
and $K_{\rm non-hol}  \supset \, \bar{\mathbf{5}}^\dag_{M} \bar{\mathbf{5}}_F, 
\,{1\over m_{\rm Pl}} \times \bigl\{
\mathbf{5}_M^\dagger\bar{\mathbf{5}}_{H,F}\mathbf{10}_F\,,\,
\bar{\mathbf{5}}_M\mathbf{5}_H^\dagger\mathbf{10}_F\,,\,
\bar{\mathbf{5}}_M^\dagger\mathbf{10}_F\mathbf{10}_F \dots \bigr\}$,
\cite{Jedamzik:2005ir}, or in the $\mathbf{16}_M + \overline{\mathbf{16}}_M$ of
SO(10) GUT, one can have operators such as
$K_{\rm hol} \supset  \overline{\mathbf{16}}_{M} {\mathbf{16}}_F, \,
\frac{1}{m_{\rm Pl}} \times \bigl\{
\mathbf{16}_{M} \mathbf{16}_F \mathbf{10}_H   \,,\,
\overline{\mathbf{16}}_{M} \mathbf{16}_F \mathbf{45}_H \dots \bigr\}$
or $K_{\rm non-hol} \supset  
\frac{1}{m_{\rm Pl}} \times \bigl\{
\mathbf{16}_{M}^\dag \mathbf{16}_F^\dag \mathbf{10}_H \,,\,
\overline{\mathbf{16}}_{M}^\dag \mathbf{16}_F \mathbf{10}_H \,,\, \dots 
\mathbf{16}_{M}^\dag \mathbf{16}_F \mathbf{45}_H \dots \bigr\}$  
\cite{Kuroda:2009}.
We note here that the supergravity features discussed above lead to an 
important difference between $K_{\rm hol}$ and $K_{\rm non-hol}$ after SUSY 
breaking: the $K_{\rm hol}$ contributions go effectively in the superpotential 
with an extra $m_{\tilde{G}}$ suppression, i.e. $W \supset m_{\tilde{G}} 
\times K_{\rm hol}$. As we will see in the following section, the above operators 
will play an important role in the cosmological fate of the LMP. 


\section{The cosmological set-up}\label{sec:cosmo}
As noted at the end of  \sref{sec:GMSB} the LMP is stable within the minimal
GMSB scenarios. If such a particle is produced at the end of inflation, 
i.e. $T_{\rm RH} \gsim M_{s_{-}}$, with $T_{\rm RH}$ the reheat temperature, then
it will typically overclose the Universe with a relic density
$\Omega_M h^2  \simeq 10^5 \left(\frac{M_{s_-}}{10^3 TeV}\right)^2$ 
\cite{Fujii:2002fv, Dimopoulos:1996gy}, unless
its mass is finely adjusted. Of course, one can avoid this 'cosmological
messenger problem' either assuming the LMP to be much lighter than 
$\sim 10^3$TeV  or that it is simply too heavy to be produced
in the early Universe. However, given our present ignorance of the actual
value of $T_{\rm RH}$ that can range from $1$MeV up to the GUT scale, 
and a rough idea about the messenger mass scale $\gsim 10^5$GeV 
\footnote{indeed, requiring the MSSM soft masses to be $\lsim 1$TeV implies
${\langle F_S \rangle \over M_X} \lsim 10^5$GeV. Furthermore,
 $M_{s-}^2 \ge 0$ imposes  $\langle F_S \rangle \le M_X^2$, thus
leading to $M_X \gsim 10^5$ GeV which gives
the mass scale of the LMP, barring fine-tuned values.} the LMP is expected
to be generically present in the very early Universe. 
As we will argue, its presence can even play an important role in making the 
gravitino a viable DM candidate \cite{Fujii:2002fv, Baltz:2001rq}, \cite{Jedamzik:2005ir,
 Lemoine:2005hu}.
 
\begin{figure}[t]
\begin{center}
\psfig{file=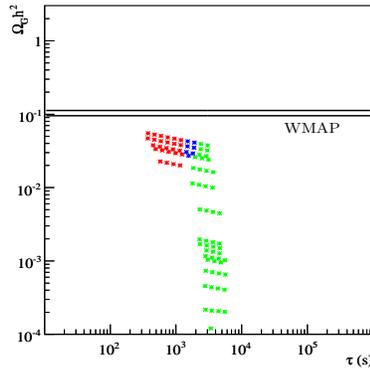,width=1.9in, height=1.9in}
\end{center}

\vspace{-.5cm}
\caption{The non-thermal stau-NLSP decay contribution $\Omega_{\tilde{G}}^{\text{non-th}} h^2$ 
to the gravitino abundance versus the NLSP lifetime, with $N_{\rm mess}= 2$,
$M_{s_{-}} = 5 \times 10^6$GeV, $\tan \beta =10$, $10 {\rm MeV} \le 
m_{\tilde{G}} \le 100 {\rm GeV}$ and a scan over $\Lambda$, 
taken from \cite{Bailly:2008these}. The horizontal band corresponds to the  
$0.095 < \Omega_{\text{CDM}}h^2 < 0.136$ WMAP consistent region. (see
\fref{fig:fig4} for the green/red color code.)}
\label{fig:fig2}
\end{figure}

\begin{figure}[t]
\begin{center}
\psfig{file=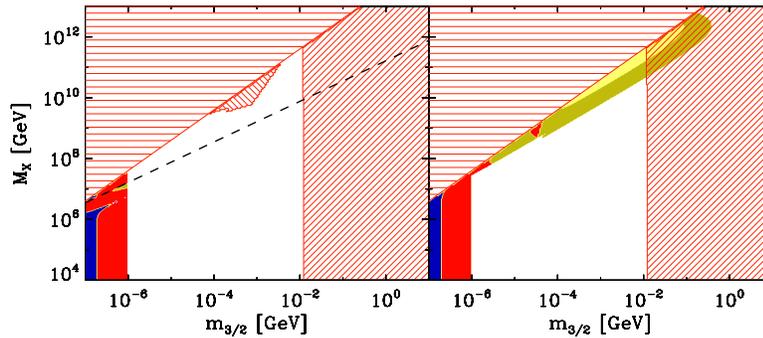,width=4in}
\end{center}
\vspace{-.5cm}
\caption{sneutrino-like LMP versus gravitino masses. The spurion is heavier (lighter) than the LMP in the 
left-hand (right-hand) panel. Green/yellow region corresponds to
gravitino cold DM with $\Omega_{\tilde{G}} h^2 < 0.3$; 
$T_{\rm RH} = 10^{12}$GeV.
(red/blue correspond to warm/hot gravitinos);
the NLSP is assumed to be a $150$ GeV Neutralino, decaying mainly 
into a photon (or a Z-boson) and a gravitino. 
The red-hatched bands to the right of each panel indicate the $m_{\tilde{G}}$ 
regions where this decay occurs after $\sim 1$ sec, thus potentially
affecting primordial nucleosynthesis.
Taken from \cite{Jedamzik:2005ir} to which we refer for further details.}
\label{fig:fig3}
\end{figure}

 In the mass range we consider,  ${\cal O}(1 \; {\rm keV}) \le m_{\tilde{G}} \le
{\cal O}(1 \; {\rm GeV})$, the gravitino is easily produced through scattering in 
the thermal bath  (see \cite{Bolz:2000fu, Pradler:2006hh, Rychkov:2007uq} and
references therein). Due to its gravitationally suppressed coupling, and in
particular that of its spin-$1 \over 2$ component which scales as
$(m_{\tilde{G}} m_{\rm Pl})^{-1}$, the leading contribution to the thermal
component of its relic density reads $\Omega_{\tilde{G}}^{\text{th}} h^2 \simeq
0.32  \left( \frac{T_{\rm RH}}{10^8 \; \text{GeV}}\right)   \left( \frac{10 \;
\text{GeV}}{m_{\tilde{G}}} \right) \left( \frac{m_{1/2}}{1 \;
\text{TeV}}\right)^2$. (Here $m_{1/2}$ denotes generically a common value of
the gaugino soft masses $M_i$.) This illustrates one of the various facets of
the so-called {\sl gravitino problem}. The dependence on $T_{\rm RH}$, a
paramter so far still poorly connected with the particle physics modelling, is
theoretically  annoying as it requires a high level of adjustment, with
basically no other observational consequences than providing an 
observationally consistent abundance for the gravitino if it is to play the 
role of DM. Perhaps more importantly, depending on the values of
$m_{\tilde{G}}$ and $m_{1/2}$ (and other parameters of the MSSM), the
gravitationally suppressed decay into (or of) the gravitino, depending on
whether it is the LSP or the next to LSP (NLSP), can equally strongly affect
the success of the  standard Big Bang nucleosynthesis (BBN) predictions; we
comment further  these issues at the end of the section. On top of
$\Omega_{\tilde{G}}^{\text{Th}}$ the gravitino abundance can have substantial
non-thermal contributions from  the decay of whatever heavier relic particles,
if such decays occur after  these particles have dropped out of thermal
equilibrium.   For instance, if only MSSM particles are present, one gets a
non-thermal contribution $\Omega_{\tilde{G}}^{\text{non-th}} h^2 = \Omega_{\rm
NLSP} h^2 \frac{m_{\tilde{G}}}{m_{\rm NLSP}}$, where $\Omega_{\rm NLSP} h^2$ is
the abundance of the essentially thermally produced NLSP which can be a
neutralino or a stau, akin to point {\sl i)} of \sref{sec:introduction}.
$\Omega_{\tilde{G}}^{\text{non-th}} h^2$ is often taken as the main source of
gravitino abundance in scenarios of gravitino DM with $m_{\tilde{G}} \gsim
150$GeV (motivated by gravity mediation) \cite{Olive:2008uf}, forgetting
altogether the uncertainties from  $\Omega_{\tilde{G}}^{\text{th}} h^2$. We
stress here that in GMSB scenarios one cannot play successfully a similar game
since, due to the lightness of the gravitino, the above 
$\Omega_{\tilde{G}}^{\text{non-th}} h^2$ cannot account alone for the
observations as illustred in \fref{fig:fig2} where the scan extends up to
$m_{\tilde{G}} = 100$GeV and the NLSP is a stau. It is then interesting to note
that for a gravitino $\gsim 1$GeV  (and $m_{\rm stau} \approx 200$GeV, a
typical configuration for a not too fine-tuned GMSB) one needs a thermal
component with $T_{\rm RH} \gsim 5 \times 10^6$ GeV in order  to reach a
suitable gravitino DM abundance. Such values of $T_{\rm RH}$ become of order
the LMP mass suggesting that the LMP (and perhaps other heavier states of the
messenger/spurion sectors) will be present in the early Universe. If so, a
different thermal history may occur, modifying the usual MSSM based estimates. 
This brings us to the crux of the scenario:  $T_{\rm RH}$ can be anywhere all
the way up to very large values. Part or all of the GMSB sectors
(\fref{fig:fig1}) are thus present early on in the thermal bath and  contribute
to the thermal production of the gravitino which is then typically  very large.
As stressed at the beginning of this section the LMP decouples  from the
thermal bath with a very large abundance causing potentially an  overcloser
problem. However the LMP is likely to decay through Planck suppressed or
gravitino suppressed operators as discussed at the end of \sref{sec:sugra}.
Such late decays occur typically after the LMP freeze-out  and would
substantially dilute the gravitino abundance through entropy  release if they
occur at a temperature where the LMP dominates the Universe {\sl and} after the
gravitino has decoupled from the thermal bath. Thus, the scenario entails the
calculation of the LMP thermal relic density yield $Y_{{}_M}$  and messenger
decay width $\Gamma_M$, and a comparison among  its freeze-out temperature
$T^f_M$, decay temperature $T_{\rm dec} \sim \Gamma_M^{1/2}$, matter domination
temperature $T_{{}_{MD}} \simeq {4 \over 3} M_{s_-} \times Y_{{}_M}$ as well as
the gravitino freeze-out temperature. One finds a substantial part of the
parameter space such that the diluted gravitino abundance is consistent with
WMAP and can represent the (cold) dark matter however large $T_{\rm RH}$ may
be!  We show an example in \fref{fig:fig3} for  $T_{\rm RH}$ as large as
$10^{12}$ GeV in the case of  $\mathbf{5}_M + \bar{\mathbf{5}}_M$ of SU(5) and with
the first operator of $K_{\rm hol} \supset \mathbf{5}_{M} \bar{\mathbf{5}}_F$ 
given in  \sref{sec:sugra} for illustration. One sees that the details of the
messenger/spurion sectors can have an important effect on the viability of the
DM scenario. For instance the small red-hatched area in the left-hand panel of
\fref{fig:fig3} corresponds to gravitino DM solutions in the scenario of
\cite{Fujii:2002fv} where  $\langle F_{S} \rangle \simeq \langle F_{\rm TOT}
\rangle$. However it corresponds to a spurion much heavier than the LMP in a
parameter space region (above the dashed black line) where  spurion mediated
LMP annihilation into gravitinos violates perturbative  unitarity, thus
theoretically unreliable. In contrast, viable solutions exist when the spurion
is lighter than the LMP, as shown by the green/yellow region on the
right-hand panel.   A systematic
study including other possible operators has  been carried out in
\cite{Jedamzik:2005ir}. A more promising case is the $\mathbf{16}_M +
\overline{\mathbf{16}}_M$ of SO(10). The LMP being an MSSM singlet in this
case, its interaction with the thermal bath is loop suppressed leading to a
much higher $Y_M$ than in the SU(5) case for comparable $M_{s_{-}}$. Taking
into account the decay induced by $K_{\rm hol}$ or $K_{\rm non-hol}$, one finds 
gravitino DM solutions when the spurion is much heavier than the LMP,
but this time 
in regions where perturbative unitarity remains reliable
\cite{Lemoine:2005hu},
\cite{Jedamzik:2005ir, Kuroda:2009}. By the same token,
one can justify here not considering explicitly the 
SBGS and messenger sectors  (left-hand part of \fref{fig:fig1}) by
assuming  them  to be much heavier than the LMP, thus playing a role
similar to that of the spurion
(i.e. essentially gravitational contributions to LMP annihilation into
gravitinos).\footnote{obviously these sectors could offer DM candidates, or lead to
cosmological closer problems on their own.
In this case they can be treated along similar lines quite symmetrical to the 
ones considered in the present study.}

 
\begin{figure}[t]
\begin{center}
\psfig{file=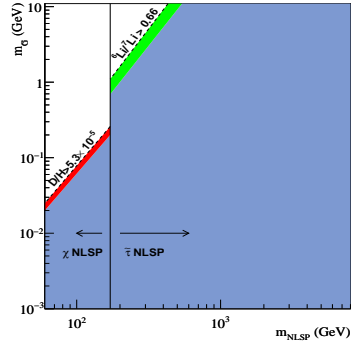,width=2in
, height=2.in
}
\end{center}
\vspace{-.5cm} 
\caption{Gravitino versus NLSP masses. same GMSB parameters as in
\fref{fig:fig2}; ${}^7$Li/H$<2.5\times 10^{-10}$ (red);
$0.015 < {}^6$Li/${}^7$Li$<0.66$ (green); SBBN: $Y_p \le 0.258$, 
$1.2 \times 10^{-5} \le$ D/H $\le 5.3 \times 10^{-5}$, 
${}^3$He/D$\le 1.72$ (light blue); 
taken from \cite{Bailly:2008yy}.}
\label{fig:fig4}
\end{figure}

Finally, let us briefly discuss the issue of primordial nucleosynthesis of 
the light elements which constitutes an important observational probe of the 
earliest epochs of the thermal history. A late decaying particle 
from physics beyond the SM (with a lifetime $\tau \sim {\cal O}(1 \, sec)$)
can affect the successful standard big bang nucleosynthesis (SBBN) 
predictions through either electromagnetic injections or
hadronically induced nuclear reactions \footnote{ see for instance \cite{Dimopoulos:1987fz}, \cite{Jedamzik:2004er}, 
\cite{Kawasaki:2004yh} and references therein and thereout.}.
This possibility has become particularly interesting in the perspective
of solving  a problematic deviation from SBBN 
of the  ${}^7$Li and  ${}^6$Li inferred observational abundancies in 
low metalicity stars. Moreover, a very efficient catalyses of the 
${}^6$Li--producing reaction can occur if the decaying particle is electrically charged 
and sufficiently long lived \cite{Pospelov:2006sc}. Constraints on physics
beyond the SM are thus of two types: conservative 
(consistency with SBBN) or speculative (solving the Lithium problems).
We illustrate these two features in \fref{fig:fig4}  
within the GMSB context \cite{Bailly:2008yy}, showing the effect of the nature 
of the NLSP on the lithium yields.

 In this respect, it is to be noted that the 
LMP decays typically at temperatures ${\cal O}(100 {\rm MeV})$ if
$M_{s_{-}} \ge 10^3$TeV, thus rendering the
gravitino DM scenarios we have described here quite safe from the BBN 
perspective. Nonetheless, one should keep in mind that 
it remains exclusively a task 
for the colliders to ultimately favor or disprove  GMSB scenarios.


\section*{Acknowledgments}
My thanks go to the organizers of DARK 2009 for the very enjoyable atmosphere 
and the quite diversified topics of the conference.
This work was supported in part by ANR under contract  
{\sc NT05-1\_43598/ANR-05-BLAN-0193-03}.

\bibliographystyle{ws-procs9x6}
\bibliography{moultaka-dark09}

\end{document}